\documentclass[fleqn,usenatbib]{mnras}

\usepackage{newtxtext,newtxmath}

\usepackage[T1]{fontenc}
\usepackage{threeparttable}
\usepackage[whole]{bxcjkjatype}
\usepackage{url}

\DeclareRobustCommand{\VAN}[3]{#2}
\let\VANthebibliography\thebibliography
\def\thebibliography{\DeclareRobustCommand{\VAN}[3]{##3}\VANthebibliography}

\usepackage{graphicx}	
\usepackage{amsmath}	

\usepackage{amssymb}	

\title[W33 in the KAGONMA survey]
{KAgoshima Galactic Object survey with Nobeyama 45-metre telescope by Mapping in Ammonia lines (KAGONMA): Star formation feedback on dense molecular gas in the W33 complex\thanks{A part of this work was carried out under the common use observation program at Nobeyama Radio Observatory (NRO).} \\ 
}

\author[T. Murase et al.]{
Takeru Murase,$^{1}$\thanks{E-mail:takerun.charvel@gmail.com}
Toshihiro Handa,$^{1,2}$
Yushi Hirata,$^{1}$
Toshihiro Omodaka,$^{1,2}$\newauthor
Makoto Nakano,$^{3}$
Kazuyoshi Sunada,$^{4}$
Yoshito Shimajiri,$^{1,5,6}$
Junya Nishi$^{1}$
\\
$^{1}$Department of Physics and Astronomy, Graduate School of Science and Engineering, Kagoshima University, 1-21-35 K\^{o}rimoto,\\ Kagoshima, Kagoshima 890-0065, Japan\\
$^{2}$Amanogawa Galaxy Astronomy Research centre, Kagoshima University, 1-21-35 K\^{o}rimoto, Kagoshima, Kagoshima 890-0065, Japan\\
$^{3}$Faculty of Science and Technology, Oita University, 700 Dannoharu, Oita, Oita 870-1192, Japan\\
$^{4}$Mizusawa VLBI observatory, NAOJ 2-12, Hoshigaoka, Mizusawa, Oshu, Iwate 023-0861, Japan\\
$^{5}$National Astronomical Observatory of Japan, Osawa 2-21-1,Mitaka, Tokyo 181-8588, Japan\\
$^{6}$ Laboratoire d’Astrophysique (AIM), CEA/DRF, CNRS, Universit\'{e} Paris-Saclay, Universit\'{e} Paris Diderot, Sorbonne Paris Cit\'{e}, 91191\\ Gif-sur-Yvette, France
}

\date{Accepted XXX. Received YYY; in original form ZZZ}

\pubyear{2021}

\begin{document}
\label{firstpage}
\pagerange{\pageref{firstpage}--\pageref{lastpage}}
\maketitle

\begin{abstract}
We present the results of NH$_3$ (1,1), (2,2) and (3,3) and H$_2$O maser simultaneous mapping observations toward the high-mass star-forming region W33 with the Nobeyama 45-m radio telescope. W33 has six dust clumps and one of which, W33 Main, is associated with a compact H{\scriptsize II} region. To investigate star-forming feedback activity on its surroundings, the spatial distribution of the physical parameters was established. The distribution of the rotational temperature shows a systematic change from west to east in our observed region. The high-temperature region obtained in the region near W33 Main is consistent with interaction between the compact H{\scriptsize II} region and the periphery molecular gas. The size of the interaction area is estimated to be approximately 1.25 pc. NH$_3$ absorption features are detected toward the centre of the H{\scriptsize II} region. Interestingly, the absorption feature was detected only in the NH$_3$ (1,1) and (2,2) transitions, with no absorption feature seen in the (3,3) transition. These complex profiles in NH$_3$ are difficult to explain by a simple model and may suggest that the gas distribution around the H{\scriptsize II} region is highly complicated.
\end{abstract}

\begin{keywords}
Stars: formation -- ISM: molecules -- ISM: H{\scriptsize II} regions
\end{keywords}



\section{Introduction}

High-mass stars ($> 8 M_\odot$) affect the surrounding environment through expansion of the H{\scriptsize II} regions, powerful outflows, strong stellar winds, and large amounts of radiation. Consequently they ultimately play a key role in the evolution of the host galaxy \citep{kennicutt2005}. In addition, the feedback from the high-mass stars influences subsequent star formation. For example, feedback from high-mass stars causes strong shocks in the surrounding molecular gas, which compresses the gas and triggers star formation \citep[e.g.][]{urquhart07,shimajiri08,Thompson12,Deharveng15,Duronea17,Paron2021}. In other cases, feedback heats the surrounding molecular gas and suppresses the fragmentation of cores or filaments. This is thought to be a contributing factor in the formation of high-mass stars \citep[e.g.][]{Bate09,Hennebelle11,Deharveng12,bate15,Hennebelle20}. In this study, we focus on the effects of the formation processes of high-mass stars on the molecular cloud environment.

Several recent observations as part of the James Clerk Maxwell Telescope (JCMT) Gould Belt survey \citep{Ward-Thompson07} have quantified the effect of radiative feedback from OB stars on molecular clouds. Using the dust colour temperature derived from the flux ratios of 450 $\micron$ and 850 $\micron$ continuum emission, they found that the dust temperature around OB stars can rise to 40 K, with effects on scales of several parsec \citep{Rumble21}. They reported that the heating of the dust may raise the jeans mass and enhance the stability of the cores of filaments against gravitational collapse \citep{Hatchell13,Rumble15,Rumble16,Rumble21}. In addition to dust data, the gas temperature map using molecular lines (e.g. CO, NH$_3$, N$_2$H$^+$, etc.) can be obtained. Large surveys of nearby star-forming regions in NH$_3$ revealed that the average gas temperature in molecular clouds with inactive star formation is around 15 K, while active star-forming cores have temperatures above 20 K \citep[e.g.][]{urquhart2015,Friesen17,hogge18,billington19,keown19,Tursun20}. In the range of hydrogen number densities $n\mathrm{(H_2)} \ga 10^{4}$ cm$^{-3}$, the dust and gas temperatures are expected to be well coupled due to frequent collision of dust grains and gas \citep[e.g.][]{Goldsmith01,Seifried17}.

Most stars are recognised to form in cluster mode \citep{Lada03,Krumholz19}. The heating of molecular gas by radiation feedback from high-mass stars in the cluster may impact the types of stars formed \citep[e.g.][]{Bate09,Rumble15}. In order to study the impact of radiative feedback, it is important to investigate the effects and scales of star-forming regions of various evolutionary stages on the surrounding environment.

There are several techniques to measure the temperature of the interstellar medium (ISM). The dust temperature $T_\mathrm{dust}$ can be estimated by fitting a single-temperature graybody model to the observed spectral energy distribution (SED) of the dust continuum thermal emission \citep{Hildebrand83}. However, there are uncertainties in the assumptions of the dust emissivity coefficient $\kappa$, and dust emissivity index $\beta$, which affect the accuracy of the derived dust temperature. CO emission lines are commonly used for molecular gas observations. In particular, the gas excitation temperature can be easily obtained from the brightness temperature of $^{12}$CO (J=1--0) by assuming the opacity thick and filling factor in the observed beam is unity. While it should be noted that the physical parameters in the centre of high-density cores may not reflect this due to optical thickness and the freezing out onto dust grains \citep[e.g.][]{Willacy98,Tafalla02,Christie12,Feng20}.

NH$_3$ has long been recognised as a good thermometer for the ISM \citep{ho83} and its line observations have the advantage of the ability to derive physical parameters such as column density and optical depth from the splitting of the inversion transition into hyperfine structure lines with only reasonable assumptions that the main transitions of the molecule are emitted under similar excitation conditions. The rotational temperature can also be estimated from the relationship between the intensity ratio of two different inversion transition lines and the optical depth. In addition, the inversion transitions in the lowest metastable rotational energy levels are easily excited in the typical temperatures of molecular clouds. Moreover, NH$_3$ molecules are abundant in the gas phase in cold and high-density environments \citep[e.g.][]{bergin97,Tafalla02}.

Previous observational studies in NH$_3$ lines have predominantly been single beam pointings toward infrared dark clouds (IRDCs), young stellar objects (YSOs) and the centres of H{\scriptsize II} regions \citep[e.g.][]{wilson1982,Rosolowsky2008,Dunham2010,urquhart11,Wienen18}, and mappings at scales of few parsec \citep[e.g.][]{keto1987,mangum92,Toujima2011,chibeze2013,urquhart2015,nakano17,billington19,burns19}. These studies have related the physical conditions of active star-forming cores to the surrounding environment. Recently, large surveys with the Green Bank Telescope (GBT) have been conducted to study the relationship between the kinematics of dense gas and star-formation in entire molecular clouds. These observations cover giant molecular clouds \citep[K-band Focal Plane Array (KFPA) Examinations of Young STellar Object Natal Environments (KEYSTONE);][]{keown19}, the Gould Belt star-forming regions \citep[the Green Bank Ammonia Survey (GAS);][]{Friesen17} and the galactic plane \citep[the Radio Ammonia Mid-Plane Survey (RAMPS); covering $10 \degr \leq l \leq 40\degr$, $\mid b \mid \leq 0 \degr .5$;][]{hogge18}. These large surveys were made with the On-The-Fly (OTF) mapping mode. However, these unbiased surveys are rather shallow, since NH$_3$ line observations require a large amount of time to detect weaker emission. Therefore, we conducted a high-sensitivity NH$_3$ imaging survey targeting dense molecular cores and the regions around them.

For our survey, we identified dense molecular cores based on the C$^{18}$O (J=1--0) imaging data obtained as a part of FUGIN \citep[FOREST Unbiased Galactic plane Imaging survey with the Nobeyama 45-m telescope;][]{umemoto17}. 
By using the {\it{Clumpfind}} algorithm \citep{Williams1994}, we catalogued 72 molecular cores / clumps in part of 1st and 3rd quadrants of the Galactic plane (i.e. $10 \degr \leq l \leq 50 \degr$, and $198 \degr \leq l \leq 236 \degr$, $\mid b \mid \leq 1 \degr$). This catalogue includes Infrared Dark Clouds (IRDCs), high-mass star-forming regions and H{\scriptsize II} regions. 
We have already finished the mapping observations towards 7 cores or clumps, which are listed in Table \ref{tab:KagonmaCores}. This paper is the first report of the KAGONMA survey project, which is an acronym of Kagoshima Galactic Object survey with Nobeyama 45-metre telescope by Mapping in Ammonia lines. We present the results of the W33 high-mass star-forming region, which is identified as KAGONMA 64.

\begin{table}
    \centering
    \caption{\label{tab:KagonmaCores} A list of KAGONMA sources which have already been observed. }
    \begin{threeparttable}
    \scalebox{0.7}{
        \begin{tabular}{ccccc}
            \hline
            KAGONMA name &
            \begin{tabular}[c]{@{}c@{}}$l$\\ {[}degree{]}\end{tabular} &
            \begin{tabular}[c]{@{}c@{}}$b$\\ {[}degree{]}\end{tabular} &
            \begin{tabular}[c]{@{}c@{}}$v_\mathrm{LSR}$(C$^{18}$O) \tnote{a}\\ {[}km s$^{-1}${]}\end{tabular}　&
            associated object \\ \hline \hline
            KAG1 & 44.312 & 0.039 & + 56.9 & G044.3103+00.0416\tnote{*}\\ 
            KAG35 & 14.613 & -0.565 & + 18.5 & G14.628-0.572\\ 
            KAG39 & 14.565 & -0.603 & + 18.7 & G14.555-0.606\\ 
            KAG45 & 14.454 & -0.102 & + 40.4 & G014.481-00.109\tnote{*}\\ 
            KAG64 & 12.798 & -0.202 & + 35.6 & W33 Main \tnote{*}\\ 
            KAG71 & 224.274 & -0.833 & + 18.0 & CMa OB1\\ 
            KAG72 & 201.446 & 0.638 & + 6.5 & G201.44+00.65\\  \hline
        \end{tabular}
        }
        \begin{tablenotes}\footnotesize
        \item[a] \cite{umemoto17}
        \item[*] H{\scriptsize II} region
        \end{tablenotes}
    \end{threeparttable}
\end{table}

\begin{table*}
    \centering
    \caption{\label{table:W33} Evolutionary stages of dust clumps.}
    \begin{threeparttable}
        \begin{tabular}{ccccc}
            \hline
            Source &
            \begin{tabular}[c]{@{}c@{}}$l$\\ {[}degree{]}\end{tabular} &
            \begin{tabular}[c]{@{}c@{}}$b$\\ {[}degree{]}\end{tabular} &
            \begin{tabular}[c]{@{}c@{}}$T_\mathrm{ex}$(C$^{18}$O) \tnote{a}\\ {[}K{]}\end{tabular} &
            Evolution Stage \tnote{b}\\ \hline \hline
            W33 A1 & 12.857 & -0.273 & 18 & High-mass protostellar object \\
            W33 B1 & 12.719 & -0.217 & 23 & High-mass protostellar object \\
            W33 Main1 & 12.852 & -0.225 & 19 & High-mass protostellar object \\
            W33 A & 12.907 & -0.259 & 18 & Hot core \\
            W33 B & 12.679 & -0.182 & 17 & Hot core \\
            W33 Main & 12.804 & -0.200 & 34 & compact H{\scriptsize II} region \\ \hline
        \end{tabular}
        \begin{tablenotes}\footnotesize
        \item[a] \cite{kohno18}
        \item[b] \cite{haschick83,immer14}
        \end{tablenotes}
    \end{threeparttable}
\end{table*}

\begin{figure*}
	\includegraphics[width=\linewidth]{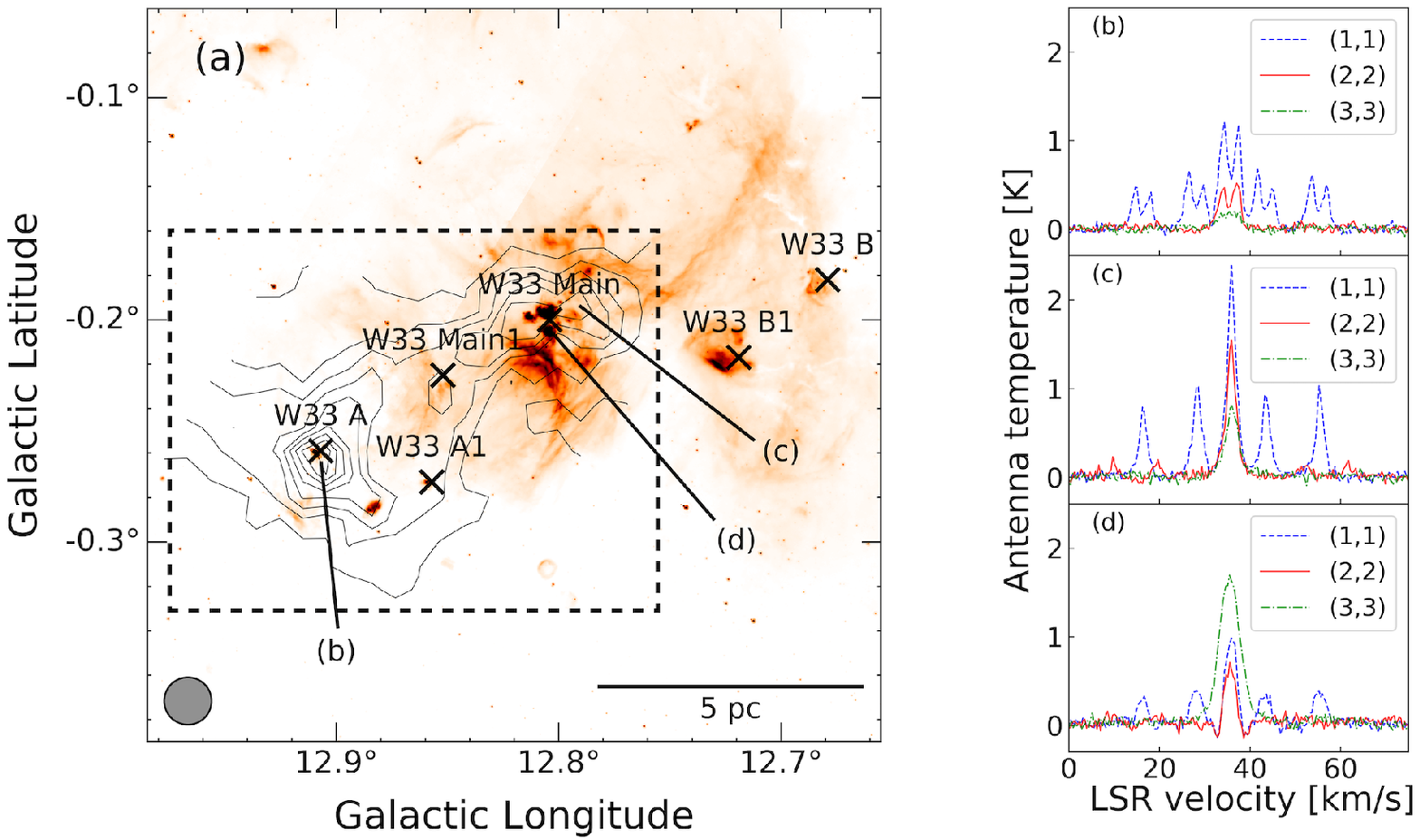}
    \caption{(a): The NH$_3$(1,1) integrated intensity map of the W33 complex in contours over the {\it Spitzer} - GLIMPSE 8.0\micron\ image. The lowest contour and contour steps are 0.8 K km s$^{-1}$ and 1.2 K km s$^{-1}$ (30$\mathrm{\sigma}$), respectively. The cross marks indicate the dust clumps reported by the ATLASGAL 870\micron\ survey \citep[e.g.][]{contreras13,urquhart14}. The dashed rectangle shows our observed area. The NRO -45m beamsize (FWHM) is indicated by the grey circle shown in the lower-left corner in Panel (a). The (b) - (d) labels indicate the positions for the profiles of panels (b)-(d). 
    (b) to (d): The profiles at the position assigned in the map with the same labels. 
    }

    \label{fig:obs_map}
\end{figure*}

Figure \ref{fig:obs_map} shows the \textit{Spitzer} -GLIMPSE 8.0$\micron$ \citep{Benjamin_2003} image of the W33 region. W33 has six dust clumps defined in the ATLASGAL (Atacama Pathfinder Experiment (APEX) Telescope Large Area Survey of the GALaxy) 870 $\micron$ survey \citep{schuller09,contreras13,urquhart14}, which are W33 Main, W33 A, W33 B, W33 Main 1, W33 A1, and W33 B1 (see Figure \ref{fig:obs_map}). \cite{immer14} reported that these 6 dust clumps are at various stages (High-mass protostellar object, Hot core, compact H{\scriptsize II} region) in star-forming processes based on their spectral energy distributions (SEDs) from centimetre to far-infrared. In Table \ref{table:W33}, the evolutionary stage of each dust clump is listed in order of earliest to latest. W33 Main harbours a compact H{\scriptsize II} region found by radio continuum observations \citep{ho83}, indicating massive star formation. Water and methanol maser emission has been detected in W33 A, W33 Main and W33 B \citep[i.e.][]{haschick90,menten91,immer13}, and OH maser sources have been detected in W33 A and W33 B \cite[i.e.][]{caswell98,colom15}. 

The distance to the W33 complex, based on annual parallax, was established as 2.4 kpc using VLBI water maser observations \citep{immer13}. W33 is located in the Scutum spiral arm of the Milky Way. Some CO line observations covering the entire W33 region were conducted \citep[e.g.][]{stier84,sridharan02,kohno18,Liu2021}. \cite{kohno18} reported W33 A, W33 Main, and W33 B1 are at a radial velocity of $\sim$ 35 km s$^{-1}$ and W33 B has a velocity of $\sim$ 58 km s$^{-1}$, while \cite{immer13} reported that these clumps exist within a single molecular cloud because these clumps have the same parallactic distance.

This paper is organised as follows: in Section \ref{sec:obs} we describe the set-up of our observations and data reduction. In Section \ref{sec:result} we present the results and estimated physical parameters of the observed area. We evaluate the influence of star formation feedback based on the rotational temperature distribution of NH$_3$ lines in Section \ref{sec:discussion}. In Section \ref{sec:conclusions}, we summarise our results and our conclusions.

\section{Observations}
\label{sec:obs}

\subsection{NH$_3$ and H$_2$O maser observations}
We made mapping observations covering a 12\arcmin $\times$ 12\arcmin\ area including W33 A and W33 Main with the Nobeyama 45-m radio telescope from 2016 December to 2019 April. We observed the NH$_3$ (J,K) = (1,1), (2,2), (3,3) and H$_2$O maser lines simultaneously. From February 2019, we observed NH$_3$ (3,3), (4,4), (5,5) and (6,6) lines at positions where the (3,3) emission line was detected (> 20 $\mathrm{\sigma}$). The (3,3) emission line was observed again for relative calibration. We used the H22 receiver, which is a cooled HEMT receiver, and the SAM45 (Spectral Analysis Machine for the 45-m telescope: \cite{kuno2011}), which is a digital spectrometer to observe both polarisations for each line simultaneously. The bandwidth and spectral resolution were 62.5 MHz and 15.26 kHz, respectively. At the frequency of NH$_3$, these correspond to a velocity coverage and resolution of 400 km s$^{-1}$ and 0.19 km s$^{-1}$, respectively. The telescope beam size was 75\arcsec\ at 23 GHz, which corresponds to 0.87 pc at 2.4 kpc. The pointing accuracy was checked every hour using a known H$_2$O maser source, M16A at ($\alpha, \delta$)$_{J2000}$=($\mathrm{18^{h}15^{m}19^{s}.4, -13\degr46\arcmin30\arcsec.0}$), and was better than 5\arcsec. The map centre was $(l,b) = (12.\degr 820,\ -0.\degr194)$. The OFF reference position was taken at $(l,b) = (13.\degr481,+0.\degr314)$, where neither C$^{18}$O (J=1-0), NH$_3$ nor H$_2$O maser was detected.

We observed 280 positions using a 37\arcsec.5 grid in the equatorial coordinates using the position switch method. For efficient observation, three ON positions were set for each OFF position, and integrations were repeated for 20 seconds at each position. To obtain a uniform map noise, the scans were integrated until the rms noise level for each polarisation of each observed line was reached to below 0.075 K.
The typical system noise temperature, $T_\mathrm{sys}$, was between 100 and 300 K. The antenna temperature, $T_\mathrm{a}^*$, was calibrated by the chopper wheel method \citep{kutner81}. We summarise the parameters for the NH$_3$ and H$_2$O maser line observations in Table \ref{table:obs_lines}.

\begin{table}
	\centering
	\caption{Transition frequencies and excitation temperatures.}
	\begin{threeparttable}
	\label{table:obs_lines}
    \begin{tabular}{ccc}
        \hline
        Transition &
        \begin{tabular}[c]{@{}c@{}}Frequency\tnote{a}\\ {[}GHz{]}\end{tabular} & 
        \begin{tabular}[c]{@{}c@{}}$E_\mathrm{u}/k_\mathrm{B}$\tnote{a}\\ {[}K{]}\end{tabular} \\ \hline \hline
        H$_2$O 6$_{12}$-5$_{12}$ (maser) & 22.235080 & - \\ \hline
        NH$_3$ (1,1)   & 23.694495  & 23.3    \\
        NH$_3$ (2,2)   & 23.722633  & 64.4    \\
        NH$_3$ (3,3)   & 23.870129  & 123.5   \\
        NH$_3$ (4,4)   & 24.139416  & 200.5   \\
        NH$_3$ (5,5)   & 24.532988  & 295.6   \\
        NH$_3$ (6,6)   & 25.056025  & 408.1   \\ \hline
    \end{tabular}
    \begin{tablenotes}
        \item[a] From the JPL Sub-millimetre, Millimetre, and Microwave Spectral Line catalogue \citep{pickett98}. $E_\mathrm{u}$ is the energy of the upper level above the ground.
    \end{tablenotes}
    \end{threeparttable}
\end{table}

\begin{figure*}
	\includegraphics[width=\linewidth]{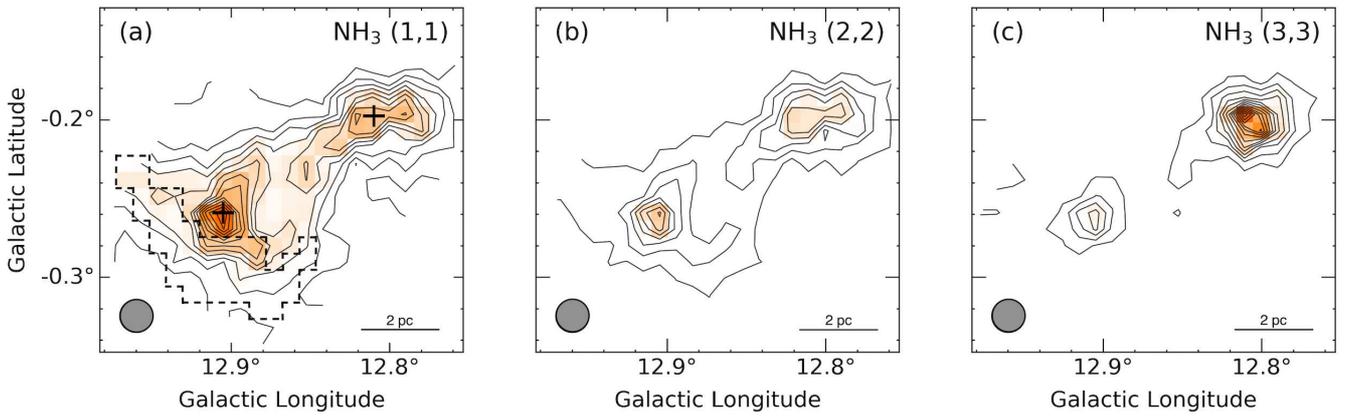}
    \caption{(a): Integrated intensity map of NH$_3$(1,1). (b): in (2,2). (c): in (3,3). The NRO45 beamsize is indicated by the grey circle shown in the lower-left corner of each panel. The lowest contour and contour steps are 20 $\sigma$ (0.8 K km s$^{-1}$) in $T_a^*$, respectively. Plus marks indicate the positions of H$_2$O maser emission. The black dashed enclosure shows the region where double-peak profiles were detected.} 
    \label{fig:IT_map}
\end{figure*}
\begin{figure*}
	\includegraphics[width=\linewidth]{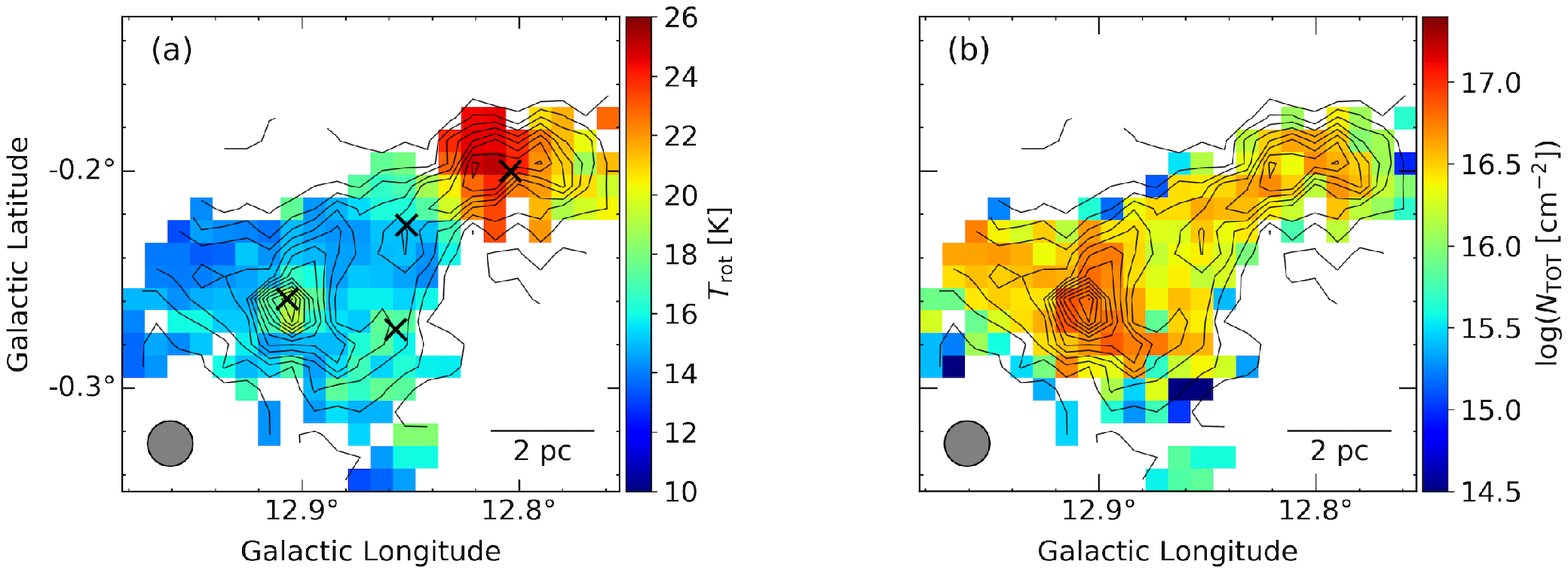}
    \caption{Spatial distributions of the physical parameters described in Section \ref{sec:deriv_physi}. The plotted values are representative along the velocity axis. (a): the rotational temperature. (b) : the total column density of NH$_3$ gas. Contours indicate the NH$_3$(1,1) integrated intensity map which are the same as in Figure \ref{fig:IT_map}-(a). The NRO45 beamsize is indicated by the grey circle shown in the lower-left corner of each Panel. Cross marks are the same as inside the dashed rectangle in Figure \ref{fig:obs_map}-(a).
    }
    \label{fig:phy_param}
\end{figure*}
\subsection{Data reduction}
\label{datareduction}

For data reduction, we use the java NEWSTAR software package developed by the Nobeyama Radio Observatory (NRO). Baseline subtraction was conducted individually for all spectra using a line function established using emission-free channels. By combining dual circular polarisations, the $rms$ noise level was typically 0.04 K at each position. A conversion factor of 2.6 Jy K$^{-1}$ was used to convert the antenna temperature to flux density.

In this paper, the intensities are presented as antenna temperature in kelvins. The NH$_3$ lines have five hyper-fine components consisting of one main line and four satellite lines. In our observations, these satellite lines were detected only in the (1,1) transition. 
The number of map positions in which $\geq 3 \sigma$ detections were achieved ub all (1,1), (2,2) and (3,3) lines were 260, 231, and 172, respectively. No emission from transitions higher than (3,3) was detected.

The NH$_3$ profiles obtained in our observations can be categorised into three types (Figure \ref{fig:obs_map} (b)-(d)). Figure \ref{fig:obs_map}-(b) shows double peak profiles detected around W33 A. 
The double-peak profiles were detected of 46 positions (The enclosure in Figure \ref{fig:IT_map}-(a)). The single peak profiles shown in Figure \ref{fig:obs_map}-(c) are typical NH$_3$ profiles. 
The intensity of these two types of profiles becomes weaker with higher excitation, while the (3,3) was detected more strongly than low excitation lines of the centre of W33 Main, where a compact H{\scriptsize II} region is located (Figure \ref{fig:obs_map}-(d)). In this region, we found absorption features at 33 km s$^{-1}$ and 39 km $^{-1}$ in (1,1) and (2,2) lines (see Section \ref{sec:abs_lines} for detail).

\section{Results}
\label{sec:result}

\subsection{Spatial distribution of NH$_3$ emission}
Figure \ref{fig:IT_map} shows the integrated intensity maps of the (1,1), (2,2) and (3,3) lines in our observed region. The velocity range of each map is between 32.0 and 40.0 km s$^{-1}$. NH$_3$(1,1) emission is extended over a region of 12\arcmin $\times$ 12\arcmin, or 10 $\times$ 10 pc at 2.4 kpc. Two NH$_3$ clumps were detected at W33 Main $(l,b) = (12.\degr 804,\ -0.\degr 200)$ and W33 A $(l,b) = (12.\degr 907,\ -0.\degr 259)$.

Although maps of both the NH$_3$ (1,1) and (2,2) line show two peaks in W33 Main with separation of about 2\arcmin, the (3,3) map shows the single peak between them. After checking the profiles of the (1,1) and (2,2) lines there, we found hints of an absorption signature. Therefore, the actual column density structure of W33 Main has only a single clump with an apparent gap inside if caused by the absorption of emission at the midway point.
We will further discuss the absorption feature in Section \ref{sec:abs_lines}.


\subsection{Linewidth correlations}
\label{sec:line_width}
\begin{figure}
	\includegraphics[width=\linewidth]{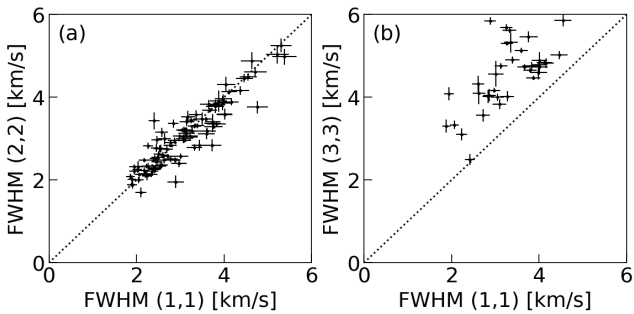}
    \caption{Scatter plots of FWHM linewidth of NH$_3$ (1,1), (2,2) and (3,3) emission. The solid lines indicates the line of equality.
    }
    \label{fig:line-width}
\end{figure}
Figure \ref{fig:line-width} shows the correlation plots of linewidths in the NH$_3$ (1,1), (2,2) and (3,3) emission lines. In Figure \ref{fig:line-width}, we only used the observed positions where single peak profiles were obtained (except for the enclosed positions in Figure \ref{fig:IT_map}-(a)).
In our observations, the range of linewidths for each NH$_3$ line was 2 to 6 km s$^{-1}$, which is broader than the expected thermal linewidth for temperatures in W33 (approximately 0.2 km s$^{-1}$ at a gas temperature of $\sim$ 20 K). These broader linewidths may be due to internal gas kinematic motions such as turbulence, outflows and stellar winds. While the linewidths of (1,1) and (2,2) emission show strong correlations, (3,3) emission tends to have systematically broader linewidths than the lower excitation transitions \citep[see also Figure 11 in][]{urquhart11}. As shown in Table \ref{table:obs_lines}, the NH$_3$ (3,3) lines requires approximately 5 times higher excitation energies than (1,1) lines. Therefore, the emission regions of higher transitions of NH$_3$ lines is considered warmer and more turbulent gas than (1,1) lines. These results are similar to single-beam observations toward the centres of MYSOs and H{\scriptsize II} regions \citep{urquhart11,Wienen18}. Our results therefore show that such linewidth correlations in NH$_3$ can also be seen on larger scales than the core scale.


\subsection{Deriving physical properties from NH$_3$ lines}
\label{sec:deriv_physi}
Using NH$_3$ line profiles we can derive several physical properties at each observed position, such as optical depth, rotational temperature and column density.

Currently, there are two major methods for deriving the optical depth and rotational temperature \citep[see][for details]{wang2020}. 
They are known as the {\it Intensity ratio} and {\it Hyper-fine fitting} methods. 
The first method is derived from the intensity ratio between two different excitation lines, assuming a Boltzmann distribution \citep[e.g.][]{ho83,mangum92}. The other method is generating a model spectrum from the radiative transfer function and searching for parameters that match the observed profiles \citep[e.g.][]{Rosolowsky2008,urquhart2015}. The common point in these methods is that the physical conditions along the velocity axis are assumed to be uniform. In general, however, molecular lines have a velocity structure, and the shape of profiles are asymmetric. Therefore, we used a method to derive physical parameters for each velocity channel based on the {\it Intensity ratio} method (see Appendix \ref{sec:CLEAN_procedure} for more details).

The equation for deriving the optical depth and rotation temperature using the {\it Intensity ratio} method is described below. The optical depth is derived from the line intensity ratios of the main and satellite lines in the (1,1) transition ~\citep{ho83}. The excitation energy differences between the hyper-fine components are very small. This allows us to assume that the main beam efficiencies, beam filling factors and excitation temperatures for all hyper-fine components are identical. Therefore, we can use
\begin{eqnarray}
\label{eq:cal_tau}
    \frac{T_a^* ( \mathrm{main} ) }{T_a^*(\mathrm{sate})} = \frac{1-\exp(-\tau)}{1-\exp(-a \tau)},
\end{eqnarray}

where the values of $a$ are 0.27778 and 0.22222 for the inner and outer satellite lines, respectively \citep{mangum92}. 

Assuming that excitation conditions of gas emitting (1,1) and (2,2) lines were the same, the rotational temperature, $T_\mathrm{rot}$ can be estimated from the intensity ratio of the (2,2) to (1,1) and optical depth at each observed position \citep{ho83} using
\begin{equation}
T_{\mathrm{rot}}(2,2;1,1) =
-41.1
\Big/
\ln \Big(
\frac{-0.282}{\tau(1,1,m)} \nonumber
\end{equation}
\begin{eqnarray}
\label{eq:trot}
\times \
\ln \left[
1-\frac{T_\mathrm{a}^*(2,2,m)}{T_\mathrm{a}^*(1,1,m)}
\times
[1-\exp(-\tau (1,1,m)])
\right]\Big),
\end{eqnarray}
where $\tau(1,1,m)$ is the optical depth of the NH$_3$ (1,1) main line.

Under the  local thermal equilibrium (LTE) condition, the column density of the NH$_3$ in the (1,1) state can be estimated using the optical depth, $\tau(1,1,m)$ and the rotational temperature, $T_{\mathrm{rot}}$ \citep{mangum92}.
\begin{eqnarray}
N(1,1) = 2.78 \times 10^{13} \tau (1,1,m) \left(\frac{T_{\mathrm{rot}}}{\mathrm{K}}\right) \left(\frac{\Delta v_{1/2}}{\mathrm{km\,s^{-1}}}\right),
\end{eqnarray}
where $\Delta v_{1/2}$ is the velocity width defined as the full width at half maxmum (FWHM) of the main line. The column densities were derived by using physical parameters at each observed position. When all energy levels are thermalized, the total column density, $N_{\mathrm{TOT}}(\mathrm{NH_3})$, can be estimated by 

\begin{equation}
N_{\mathrm{TOT}}(\mathrm{NH_3})=
N(1,1) \sum_J \sum_K 
\left(\frac{2g_\mathrm{J} g_\mathrm{I} g_\mathrm{K} }{3}
\exp
\left[
23.3-\frac{E_\mathrm{u}(J,K)}{k_\mathrm{B} T_{\mathrm{rot}}}
\right]
\right)
\end{equation}
where $k_\mathrm{B}$ is the Boltzmann constant, $g_J$ is the rotational degeneracy, $g_I$ is the nuclear spin degeneracy, $g_K$ is the K-degeneracy, and $E(J,K)$ is the energy of the inversion state above the ground state.

\subsection{Molecular gas properties}
\label{sec:molecular_gas_prop}

In this subsection, we will report the spatial distribution of the derived physical parameters in W33 (Figure \ref{fig:phy_param}), which were solved in each velocity channel. The physical parameters at each position were derived when the signal to noise ratio (S/N) of all peaks of NH$_3$ (1,1) hyper-fine components and (2,2) main line are above 3$\sigma$. The optical depth and rotational temperature errors were estimated to be $\pm$ 0.10 and $\pm$ 0.4 K, respectively (see Appendix \ref{sec:error_esti}).

The derived optical depth ranges from 1 to 2, and its mean value over all observed positions was 1.24 $\pm$ 0.10. The optical depth distribution was not significantly different at each position. 

Maps of the rotational temperature and total column density are shown in Figure \ref{fig:phy_param}. The range of rotational temperature was between 12 and 25 K (Figure \ref{fig:phy_param}-(a)). We found a clear difference between galactic east and west parts, corresponding to W33 A and W33 Main, respectively. All pixels in W33 A were colder than 18 K, and most pixels in W33 Main were warmer than 20 K.
In our NH$_3$ observations, the temperature change was a particularly noticeable characteristic. We will discuss the relationship between the molecular gas temperature and star formation feedback in Section \ref{sec:SFfeedback}.

The value of total column density ranges between $2 \times 10^{15}$ to $8 \times 10^{16}$ cm$^{-2}$ as shown in Figure \ref{fig:phy_param}-(b). The peak total column density of W33 Main and W33 A was $(5.1 \pm 0.1) \times 10^{16}$ cm$^{-2}$ and $(7.5 \pm 0.1) \times 10^{16}$ cm$^{-2}$, respectively. In contrast to the difference in the rotational temperature, there was no significant difference in the total column density between W33 Main and W 33 A. A weak dip in the measured column density at the centre of W33 main is apparently due to the compact H{\scriptsize II} region.

\subsection{H$_2$O maser detection}
In our observations, H$_2$O maser emission was detected in W33 A on the $21^{\mathrm{st}}$ May in 2017 with 26.3 Jy at $(l,b)=(12.\degr905, -0.\degr257)$ and W33 Main on the $07^{\mathrm{th}}$ May in 2016 with 10.7 Jy at $(l,b)=(12.\degr811, -0.\degr194)$. These masers are positionally consistent with those reported in the parallax observations of \cite{immer13}. H$_2$O maser emission is believed to be a signature of star formation in its early evolutionary stages \cite[e.g.][]{sunada07,urquhart11}. Therefore, the detection indicates both W33 A and W33 Main host star-forming activity.

\section{Discussion}
\label{sec:discussion}
\subsection{Velocity components in W33 complex}
\label{sec:vel_comp}
\begin{figure}
	\includegraphics[width=\linewidth]{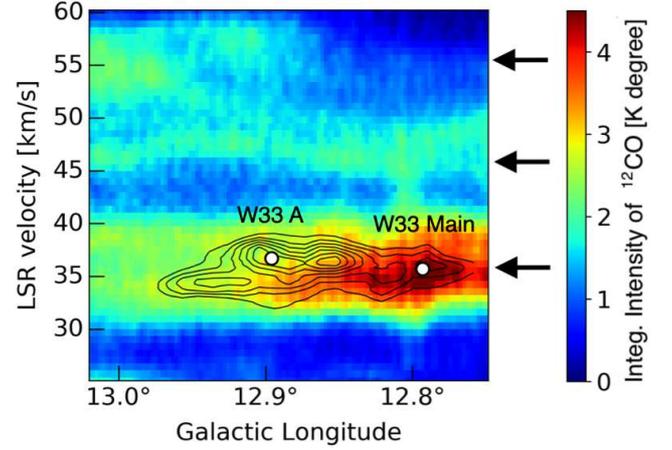}
    \caption{Longitude-velocity diagram of the $^{12}$CO (J=1-0) emission using FUGIN data (colour image) and the NH$_3$(1,1) main line emission (contour). The white circles indicate the peak velocity of W33 A and W33 Main in the C$^{18}$O(J=1-0) emission line \citep{umemoto17,kohno18}. The arrows show the three velocity components at 35 km s$^{-1}$, 45 km s$^{-1}$ and 55 km s$^{-1}$. The lowest contour level and contour intervals are 0.06 K degree and 0.03 K degree, respectively.}
    \label{fig:l_v_dirgram_CO_HN3}
\end{figure}

\cite{kohno18} and \cite{dewangan20} reported three velocity components at 35 km s$^{-1}$, 45 km s$^{-1}$ and 55 km s$^{-1}$ in the W33 region from FUGIN CO survey data \footnote{The 55 km s$^{-1}$ component is reported as 58 km s$^{-1}$ in \cite{kohno18}, and 53 km s$^{-1}$ in \cite{dewangan20}.}. The 35 km s$^{-1}$ and 55 km s$^{-1}$ velocity components exhibit similar spatial distributions \citep[see Figure 6 in][]{kohno18}. On the other hand, the 45 km s$^{-1}$ velocity component shows weak emission extended over the wider W33 complex, and its spatial distribution is not exclusively associated with the W33 complex. \cite{kohno18} concluded that the 45 km s$^{-1}$ velocity component is unrelated to the star formation activity in W33 complex. Figure \ref{fig:l_v_dirgram_CO_HN3} shows the longitude–velocity diagram using our data and FUGIN $^{12}$CO (J=1--0) data \footnote{\url{http://jvo.nao.ac.jp/portal/nobeyama/}} integrated over galactic latitudes between -0.$\degr$35 to -0.$\degr$12, where contours indicate the NH$_3$ (1,1) main line. The dominant emission in CO and NH$_3$ is detected at 35km s$^{-1}$ and gas at this velocity is considered to be a molecular cloud related to the star formation activity in W33.

\begin{figure}
	\includegraphics[width=\linewidth]{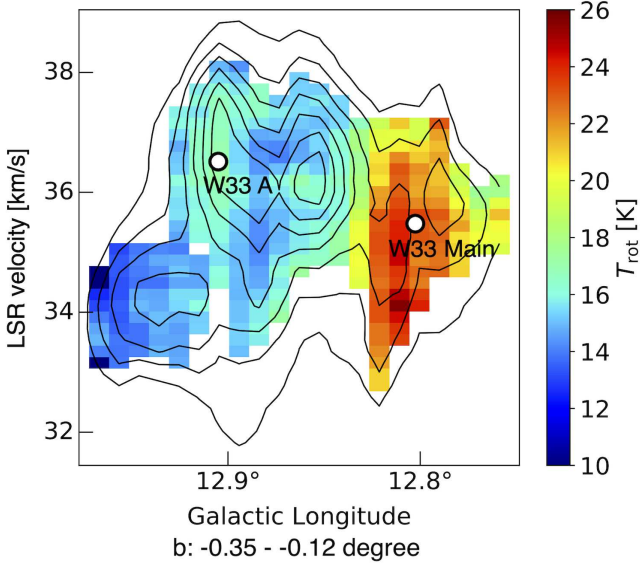}
    \caption{The longitude-velocity diagram of the rotational temperature. White circles, lowest contour level and contour interval are same as in Figure \ref{fig:l_v_dirgram_CO_HN3}.}
    \label{fig:temp_l_v}
\end{figure}

Figure \ref{fig:temp_l_v} shows the longitude–velocity diagram of the our NH$_3$(1,1)
data and rotational temperature. In the remainder of this paper, we focus on the 35 km s$^{-1}$ velocity component. Figure \ref{fig:temp_l_v} shows that there are three velocity sub-components centred on 35 km s$^{-1}$. In particular, we can find that NH$_3$ splits into two velocity components at W33 A (an easternmost component at 34.5 km s$^{-1}$, and the main component of W33 A at 36.0 km s$^{-1}$). These components with different velocities give rise to the double-peak profiles at positions shown in Figure \ref{fig:IT_map}-(a)
and have different properties as shown below. From Figure \ref{fig:obs_map}-(b), the intensity of the satellite lines are differed between the two velocity components, although the peak intensities of the NH$_3$(1,1) main line components are approximately the same. It suggests that the optical depths are different in these components, since the intensity ratio of the main and satellite lines depends on optical depth. The mean optical depth of each velocity component was 1.39 (34.5 km s$^{-1}$) and 1.50 (36.0 km s$^{-1}$), respectively.
We also investigated temperature differences between these two components. The emission at the 34.5 km s$^{-1}$ is colder than 16 K, while the 36.0 km s$^{-1}$ component is about 18 K. 
However, within each velocity component, the temperature is almost uniform (Figure \ref{fig:l_v_dirgram_CO_HN3}), suggesting there is no direct interaction between these components. In addition, there was no change in the temperature in the region between these two velocity components. We obtained no evidence of interaction such as collisions between these components in our observations. On the other hand, the NH$_3$ emission associated with W33 Main at $v_\mathrm{LSR}$= 35 km s$^{-1}$ is warmer than 20 K and will be discussed further below. 

\subsection{Star formation feedback traced by gas temperature}
\label{sec:SFfeedback}

In Section \ref{sec:molecular_gas_prop}, we show that the temperature changes more significantly
in the W33 complex than the other physical parameters. In this section, we use the temperature distribution to discuss the influence range of star formation activity.

The rotational temperature in W33 Main is higher than the other sub-components (Figures \ref{fig:phy_param}-(a) \& \ref{fig:temp_l_v}). This suggests that embedded compact H{\scriptsize II} regions are having an impact on the physical conditions of the surrounding molecular gas.
This feature of temperature distribution is also seen in some IRDCs, MYSOs and H{\scriptsize II} regions \citep{urquhart2015,billington19}. Previous studies measuring molecular gas temperature have reported that quiescent regions exhibit temperatures of 10 -- 15 K, while active star-forming regions associated with massive young stellar objects and H{\scriptsize II} regions show temperatures higher than 20 K \citep[e.g.][]{urquhart2015,Friesen17,hogge18,billington19,keown19}. In this study, the observation points that measured temperatures higher than 20 K are defined as the region under the influence of star-formation feedback.

Using Figure \ref{fig:phy_param}-(a), we estimated the size of the influenced area. The projected area showing more than 20 K was estimated from the sum of the grid points (each of $37\arcsec.5 \times 37\arcsec.5=0.19$ pc$^{2}$), resulting in a total size of 4.92 pc$^2$. Its equi-areal radius was 1.25 pc. The apparent size of the compact H{\scriptsize II} region is $12\arcsec.6 \ \times \ 4\arcsec.6$ based on the 5 GHz continuum map obtained by \cite{White2005} with the VLA, which corresponds to 0.15 pc $\times$ 0.05 pc at 2.4 kpc. The heated area is several times larger than the compact H{\scriptsize II} region.

In a previous study by our group, we investigated the size of a molecular gas cloud affected by the H{\scriptsize II} region at the edge of the Monkey Head Nebula (MHN) \citep{chibeze2013}. They reported no apparent impact of the extended H{\scriptsize II} region of the MHN. However, the molecular gas around the compact H{\scriptsize II} region S252A has higher temperatures and the size of the heating area was 0.9 pc. We can expect that there may be a relationship between the size of the heating area and the properties of the heating source, although the continuum size of S252A is unknown. In order to make more certain statements about any possible relationship, observations of more regions are required.

No temperature increase was obtained in W33 A, where the strong NH$_3$ emission was detected. In several previous studies, a large-scale outflow was reported in the centre of W33 A \citep{galvan10,kohno18}. Using $^{12}$CO(J=3--2) and (J=1--0) data, \cite{kohno18} investigated the intensity ratio, $R_\mathrm{3-2/1-0}$, for the three velocity components at 35 km s$^{-1}$, 45 km s$^{-1}$ and 55 km s$^{-1}$. A high $R_\mathrm{3-2/1-0}$ was found in W33 A and W33 Main. These results are understood to be due to the outflows from protostellar objects and heating by massive stars \citep{kohno18}. However, in our results, the temperature distribution estimated by NH$_3$ showed no evidence of heating the molecular cloud around W33 A. We consider that gas heating by outflow is not effective, or the size of the gas heated by outflows in W33 A is significantly smaller than our beam size (75\arcsec). High-resolution observations may be required to investigate the impact of such stellar feedback in more detail.

\subsection{Comparison of emission and absorption components}
\label{sec:abs_lines}

\begin{figure*}
	\includegraphics[width=\linewidth]{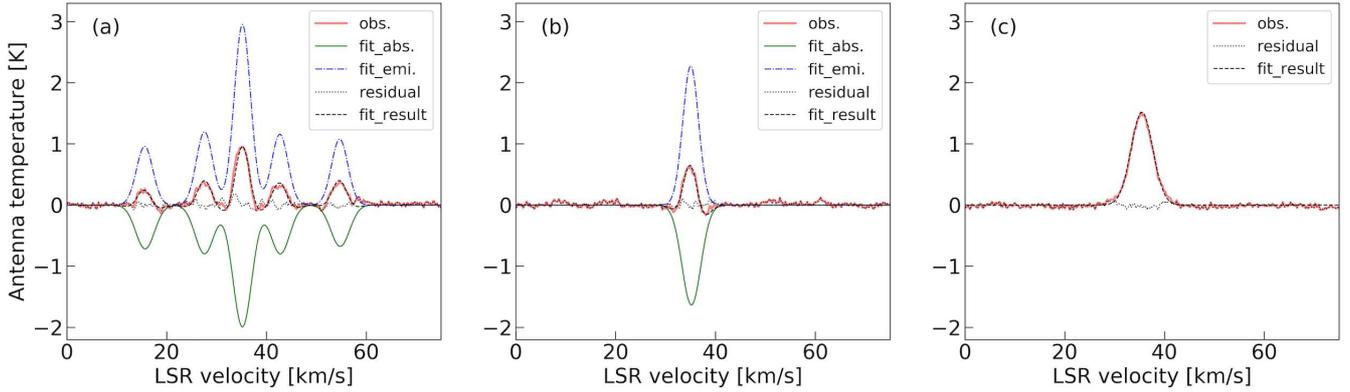}
    \caption{The profile in Figure \ref{fig:obs_map}-(d) is reproduced by a multi-component Gaussian. (a) shows the fit results of the NH$_3$(1,1) profile, (b) shows the (2,2) profile, and (c) shows the (3,3) profile. The blue dash dotted lines, green lines, red lines, and black dashed lines indicate the observed profile, absorption component, emission component, and fitting result profile, respectively. The dotted lines show the residual profile in the same colour. Since no absorption feature was found in the (3,3) profile, we apply only an emission component.
    }
    \label{fig:fit_abs}
\end{figure*}

We tried to reproduce the profile shown in Figure 1-(d) using a combination of emission and absorption using a multi-component Gaussian profile with positive and negative peaks for each of the five hyperfine lines in (1,1) and the main line in (2,2). Then we took the peak intensity, linewidth, and central velocity of each hyperfine line as free parameters. Figure \ref{fig:fit_abs} shows our result for the (1,1) - (3,3) lines. In this result, the peak velocities of both components are consistent within the error. 

High angular resolution observations toward W33 Main with the GBT, the Max Planck Institute for Radio Astronomy (MPIfR) 100 -m radio telescope and Very Large Array (VLA) also detected the absorption feature \citep{wilson1982,keto1989,urquhart11}. NH$_3$ gas in front of a bright continuum source is seen as an absorption feature \citep[e.g.][]{keto1987,henkel2008}. The GBT profile shows the absorption feature more clearly \citep[see Figure A1 of][]{urquhart11}. This suggests that the size of the continuum sources are smaller than the beamsize of the NRO 45-m. Interferometric observations of the radio continuum also support our interpretation \citep[e.g.][]{haschick83,immer14}. However, in the (3,3) line, neither observations with NRO -45m, MPIfR, nor GBT showed an absorption feature. We will discuss this further in the next subsection.

\begin{figure}
	\includegraphics[width=\linewidth]{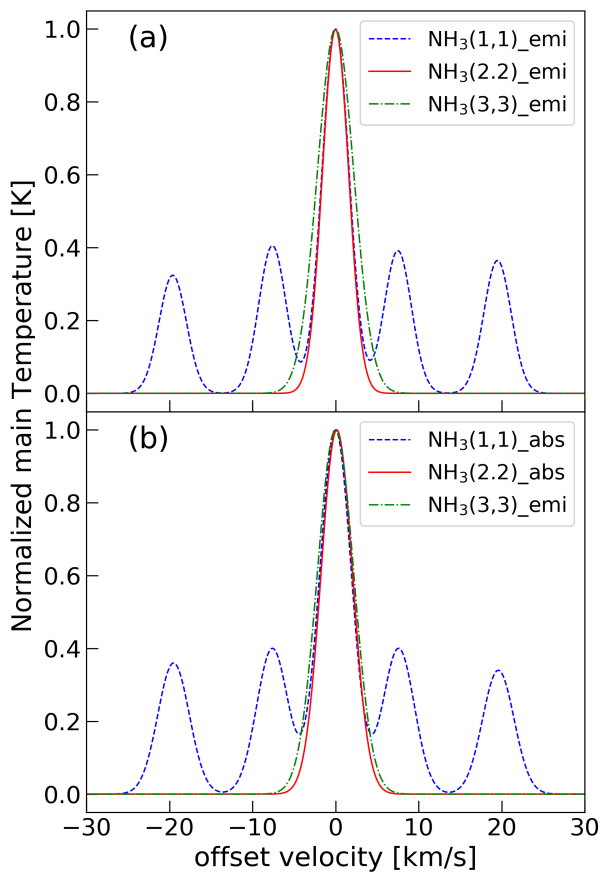}
    \caption{Comparison of the emission and absorption components obtained by Gaussian fitting. The horizontal axis indicates the offset velocity from central velocity of each transition main line. The vertical axis shows the relative intensity normalised by each main line peak. (a) shows the NH$_3$ (1,1) - (3,3) emission profiles. (b) shows the comparison between the negative-intensity profiles of the absorption components of (1,1) and (2,2), and the positive-intensity profile of (3,3).
    }
    \label{fig:comp_spectra}
\end{figure}

The decomposed line profiles show an interesting property in Figure \ref{fig:fit_abs}. Previous studies in NH$_3$ lines have reported the combination of absorption and emission lines with different peak velocities such, as a P-Cygni or inverse P-Cygni profiles, in all observed transitions \citep[e.g.][]{wilson1978,urquhart11}. In W33 Main, the peak velocity of the emission and absorption exhibited consistent velocities within their errors (see NH$_3$ (1,1) and (2,2) spectra in Figure \ref{fig:fit_abs}). We also compared the line-width of both components (Figure \ref{fig:comp_spectra}). The {\it absorption} components in (1,1) and (2,2) lines show the same line-width as the {\it emission} in the (3,3) line. Suggesting they originate in the same gas cloud. In many molecular cores, the higher transitions of NH$_3$ emission lines are thought to be emitted from more compact regions and also show broader line-widths \citep[][and see also Section \ref{sec:line_width} of this paper]{urquhart11}. Our results may indicate that the absorption component traces more turbulent gas close to the continuum sources.

Because an absorption feature delineates the physical properties of gas in front of a continuum source, it can be used to can separate the physical properties along the line of sight and within an observed beam. The estimated optical depth and rotational temperature are $\tau$ = 0.89 $\pm$ 0.10 and $T_\mathrm{rot}$ = 20.9 $\pm$ 0.4 K in the emission component, $\tau$ = 0.91 $\pm$ 0.10 and $T_\mathrm{rot}$ = 21.1 $\pm$ 0.4 K in absorption component. The physical parameters of the two components are the same within the error. This suggests that the physical conditions of NH$_3$ gas surrounding the continuum sources are the same. Therefore, the H{\scriptsize II} region located in the centre of W33 Main may still be embedded in dense molecular gas with the same motion as the surrounding environment.

\subsection{Absence of the absorption in NH$_3$ (3,3)}
An absorption feature must be located in front of a continuum background and be affected by its brightness. As mentioned in Section \ref{sec:abs_lines}, the continuum source size is smaller than the beamsize of the NRO 45-m. Previous studies have reported objects that show P-Cygni and inverse P-Cygni profiles in all inversion transitions in NH$_3$ \citep[e.g.][]{wilson1978,burns19}. In models assuming a spherically symmetric molecular gas cloud with the continuum source in its centre, these features can reveal the expansion or contraction of the gas motion in front of the continuum source. The emission and absorption features of NH$_3$ (1,1) and (2,2) lines obtained in our observations are detected with the same line-of-sight velocity. However, for the (3,3) transition, only the emission component was detected, without any hint of absorption. We tried to explain all transition profiles in our observations consistently, but failed under the spherically symmetric model after considering the following three possibilities.

An absorption future requires a bright continuum background. Therefore, we estimated the brightness temperature of the H{\scriptsize II} region in W33 Main using an electron temperature and emission measure of 7800 K and $1.5 \times 10^{6}$ pc cm$^{-6}$ from observations at 15.375 GHz with the National Radio Astronomy Observatory (NRAO) 140-foot telescope \citep{schraml69}. The expected brightness temperature of the continuum emission at our observed frequency, $T_\mathrm{cont}$, is calculated to be about 6 K which is sufficient to produce an absorption feature. Since the absorption features are detected both in (1,1) and (2,2) lines, and since NH$_3$ inversion transitions are detected in a narrow frequency range, the continuum brightness temperature is almost the same for all observed NH$_3$ lines. The value of $T_\mathrm{cont}$ at the frequency of NH$_3$ (3,3) is therefore also strong enough to produce an absorption feature. 

The spatial distribution of the NH$_3$ gas observed in the (3,3) line may differ from the other lines. If no NH$_3$ gas for producing absorption is located in front of the continuum source, we would not observe any absorption feature which contradicts the detection of absorption features at NH$_3$ (1,1) and (2,2). It is unlikely that the distributions of the same molecular species are vastly different at different excitation levels. \cite{wilson1982} observed the centre of W33 Main with the MPIfR 100-m radio telescope in the NH$_3$ (1,1), (2,2), (3,3), and (4,4) lines. They detected absorption features only in para-NH$_3$ lines \citep[see Figure 1 of][]{wilson1982}. They also failed to explain the absence of absorption in (3,3) line under LTE assumption in all four levels even when considering two molecular clouds with different temperatures. 

NH$_3$ maser emission is another possibility. The NH$_3$ (3,3) maser has been detected in star-forming regions [e.g. W51 - \cite{zhang95}; G030.7206−00.0826 - \cite{urquhart11}; G23.33−0.30 - \cite{walsh2011,hogge2019}], which has been reported to have a narrow line-width. \cite{wilson1982} proposed a similar model; only (3,3) in a state of population inversion. However, this requires coincidental masking of the absorption component by maser emission over the full velocity width, because the (3,3) emission line has the same or broader line-width than the thermal excitation line of NH$_3$ (1,1) and (2,2) lines as shown in Figure \ref{fig:comp_spectra}. Therefore, this possibility is not realistic.

High-angular resolution and high sensitivity multi transition NH$_3$ observations are required to investigate an explanatory model in more detail. It should be possible with the Square Kilometre Array (SKA) and next-generation Very Large Array (ngVLA).

\section{Conclusions}
\label{sec:conclusions}
We performed mapping observations toward the W33 high-mass star-forming region in NH$_3$ (1,1), (2,2), (3,3) and H$_2$O maser transitions using the Nobeyama 45-m radio telescope. Our observations detected only a single velocity component around 35 km s$^{-1}$. NH$_3$ (1,1) and (2,2) lines are extended over the observed region. From these observations, the distribution of the physical parameters of the dense molecular gas was obtained. Consequently, the molecular gas surrounding the H{\scriptsize II} region located at W33 Main was found to exhibit a higher temperature (> 20 K) than the rest of the observed area. The size of the influence area is estimated at approximately 1.25 pc. The heating source of the molecular gas is considered to be the compact H{\scriptsize II} region in W33 Main. Strong NH$_3$ emission was also detected in W33 A, however, no temperature increase in its molecular gas was obtained.

Molecular gas in front of the compact H{\scriptsize II} region in W33 Main was detected as an absorption feature. Gaussian fitting of the emission and absorption components reveals that the peak velocities of both components are almost the same. This suggests that the continuum source located in the centre of W33 Main may still be embedded in the dense molecular gas. Curiously, the absorption feature was detected only in the NH$_3$ (1,1) and (2,2) transitions but not in the (3,3) transition. We tried to explain these NH$_3$ profiles using spherically symmetric models but concluded that any simple model could not explain the observed profiles in all three transitions towards W33 Main. It is possible that the spatial distributions of NH$_3$ (3,3) emitted region and the continuum sources may be different. 

\section*{Acknowledgements}
We would like to thank the Nobeyama Radio Observatory staff members (NRO) for their assistance and observation support. We also thank the students of Kagoshima University for their support in the observations. The 45-m radio telescope is operated by the Nobeyama Radio Observatory, a branch of the National Astronomical Observatory of Japan. This publication makes use of data from FUGIN, FOREST Unbiased Galactic plane Imaging survey with the Nobeyama 45-m telescope, a legacy project in the Nobeyama 45-m radio telescope. This study used \textsc{astropy}, a Python package for Astronomy \citep{astropy_2013,astropy_2018}, \textsc{aplpy}, a Python open-source package for plotting \citep{aplpy2019}, \textsc{matplotlib}, a Python package for visualization \citep{Hunter_2007}, \textsc{numpy}, a Python package for scientific computing \citep{harris_2020}, and Overleaf, a collaborative tool.

\section*{Data availability}
The NH$_3$ data underlying this article will be shared on reasonable request to the corresponding author.



\bibliographystyle{mnras}
\bibliography{murase_et_al_2021} 




\appendix

\section{A technical description to estimate the physical parameters}
\label{sec:fitting_method}

\subsection{The `CLEAN' procedure for an NH$_3$ profile}
\label{sec:CLEAN_procedure}

  \begin{table*}
    \centering
	\caption{The velocity offset between the main line and four satellite lines in NH$_3$(1,1) \citep{krieger17,ho83}}
	\label{vel_off}
    \begin{tabular}{ccccccc}
        \hline
        transition & $F_1=0 \rightarrow 1$ & $F_1=2 \rightarrow 1 $ &
        \begin{tabular}[c]{@{}c@{}} $F_1=1 \rightarrow 1$ \\ $ 2 \rightarrow 2 $\end{tabular} &
        $F_1=1 \rightarrow$ 2 & $ F_1=1 \rightarrow 0$ \\ \hline \hline
        offset [MHz] & 0.92 & 0.61 & 0 & -0.61 & -0.92 \\
        offset [km s$^{-1}$] & -19.48 & -7.46 & 0 & 7.59 & 19.61\\ \hline
    \end{tabular}
\end{table*}

In many investigations using NH$_3$, the integrated or peak intensity of lines are used to derive the gas physical parameters. However, for objects with velocity structure this procedure may not be ideal because of the non-linearity of the equations in derivation. Therefore, we should estimate the physical parameters in each velocity component at first.

In this subsection, we describe our method used in this paper to derive the physical parameters based on the CLEAN algorithm. The CLEAN algorithm was devised by \cite{hogbom1974} and is the most used iterative method to improve radio interferometer images. In general, CLEAN is used in a two-dimensional map with a fixed dirty beam over the whole imaging field. In our method, we assumed that the detected intensities of each channel are a Dirac delta function for each velocity component. However, the NH$_3$ inversion transition line exhibits a hyperfine structure and the observed five-line intensities depend on the optical depth. We employ the hyperfine structure pattern with an optical depth as the dirty beam for each delta function component.

\begin{figure*}
	\includegraphics[width=0.85\linewidth]{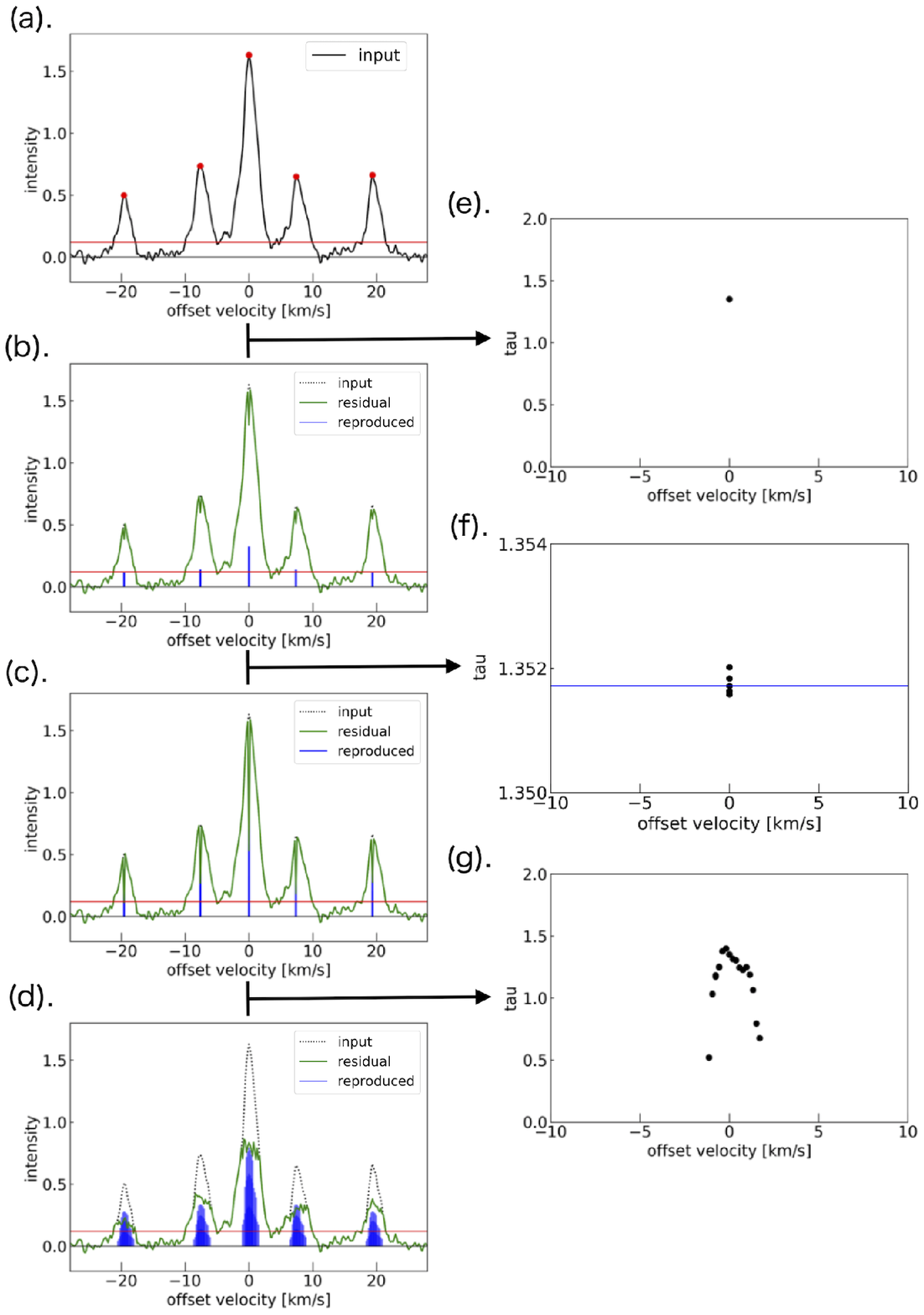}
    \caption{The workflow of our method, based on an observation position at $(l, b)$=(12.801, -0.196). Panel (a) $\sim$ (d) and (e) $\sim$ (g) show the profiles, and the estimated optical depth in each step, respectively. The red line in panel (a) $\sim$ (d) indicates the 3 $\sigma$ level. Read the main text for further detail.
    }
    \label{fig:fit_method}
\end{figure*}

Our method is composed of the following 6 steps (Figure \ref{fig:fit_method}).

\begin{enumerate}
  \item Find the peak intensity and velocity of the main line from the input profile (Figure \ref{fig:fit_method}, panel (a)). The frequency and corresponding velocity offsets between the NH$_3$(1,1) main line to the other four satellite lines are fixed as shown in Table \ref{vel_off}. This can give the five hyper-fine line intensities (indicated as red points in panel (a)). From this dataset, the four intensity ratios between the main line and satellite lines are obtained.

  \item Estimate the optical depth using each intensity ratio following equation \ref{eq:cal_tau}. To get the optical depth for a single velocity component, we take the median of the estimated four optical depths. (black dots in panel (e)). 
  
  \item Using the estimated optical depth and the main line intensity after step (ii) we reproduce the intensity of each satellite line.
  
  \item Subtract the scaled reproduced main and satellite lines from the original profile. We call the scaling factor of the reproduction as a {\it{gain factor}} $G < 1$ (indicated as blue bars in panel (b)). This {\it{gain factor}} corresponds to the {\it{loop gain}} in the CLEAN algorithm. In this paper, we adopted a {\it{gain factor}} of 0.2. 
  
  \item The steps (ii) through (iv) are iterated until one of the five hyper-fine intensities fall below a 3 $\sigma$ noise level (panel (c)). In one velocity channel, it is typically iterated 3 to 5 times. The resultant profile is used as the {\it{input}} profile in step (i). (blue horizontal line in panel (f) shows the median value of the optical depth for a channel).
  
  \item The steps from (i) to (v) are iterated $n$ times for the subtracted profile. We recommend limiting the velocity range to the main line and searching for the peak intensity (which corresponds to the {\it {cleanbox}} in the CLEAN). These steps (i) to (v) continue until the peak intensity of the residual main line is below a 3 $\sigma$ noise level (panel (d)).
  

\end{enumerate}

Figure \ref{fig:fit_method}, panels (d) and (g) show the final resulting profile and optical depth of each velocity channel. Using the optical depths and the intensity ratio of NH$_3$ (1,1) and (2,2) at each velocity channel, the rotational temperature can be estimated.

\subsection{Error Estimation by the Monte Carlo method}
\label{sec:error_esti}
\begin{figure}
	\includegraphics[width=\linewidth]{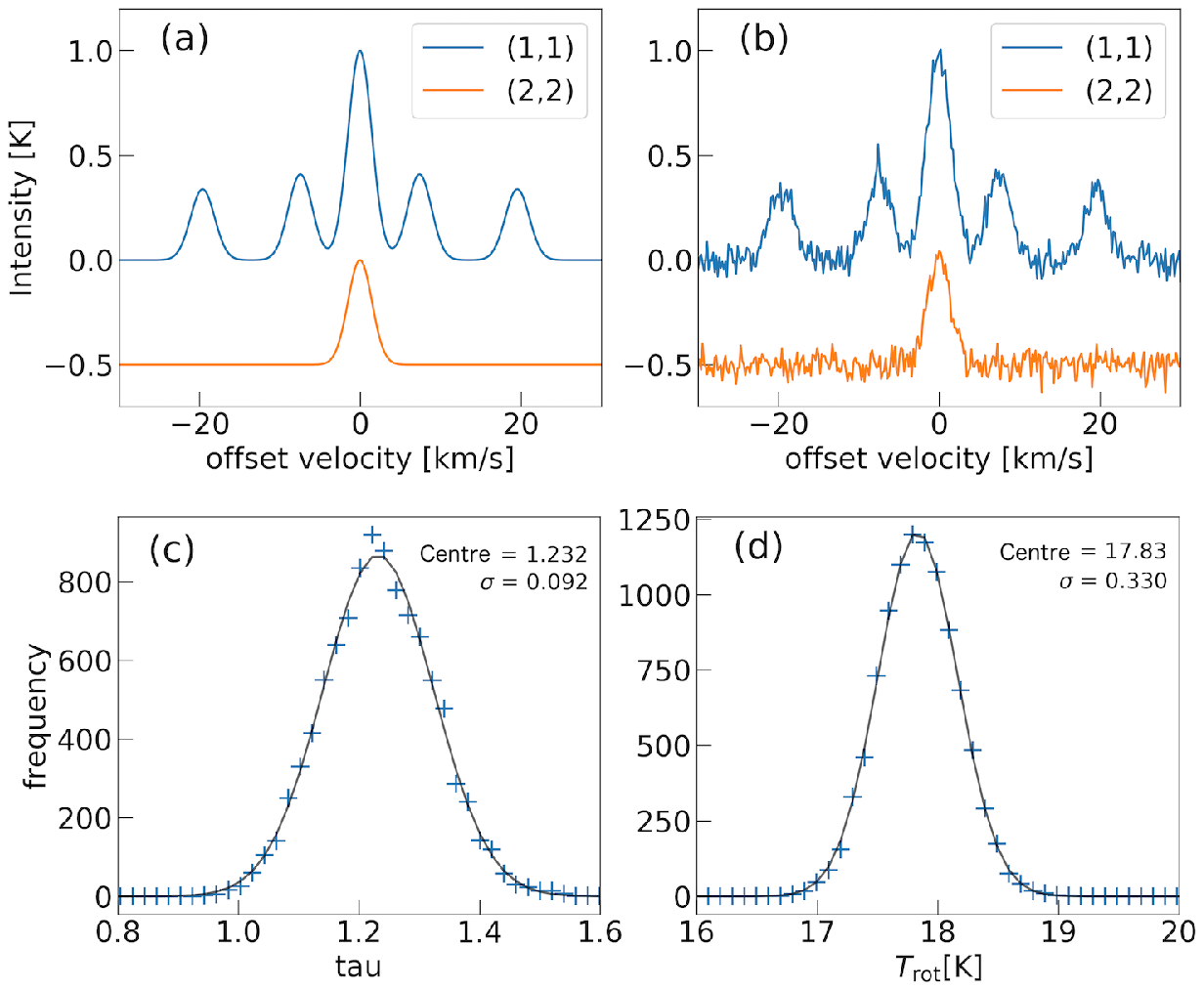}
    \caption{The result of error estimate in our method. (a) shows the model profile of NH$_3$ (1,1) (blue) and (2,2) (orange). (b) shows profiles with Gaussian noise (rms 0.04 K) added to the model profile shown in panel (a). Panel (c) and (d) indicates the frequency distributions of the optical depth and rotational temperature, respectively. Black line shows the best-fit log-normal function.
    }
    \label{fig:error_esti}
\end{figure}

Evaluating the error of observed parameter estimation is important. Although the error is straightforward when calculating the value from direct measurement, an indirect parameter derived through non-linear equations is complicated. Instead of non-linear error propagation analysis, we used a Monte Carlo method in this paper. For the sample data of NH$_3$ (1,1), we adopt a noise-free Gaussian profile assuming a constant optical depth in the line-of-sight direction. Here, we used an optical depth value of 1.24 (the average value across the W33 complex). The peak intensity of the (1,1) main line was 1.0  (Figure \ref{fig:error_esti}-(a)). For the (2,2) data, we used a Gaussian profile with half the intensity of the (1,1) emission. Using equation \ref{eq:trot}, the value of the rotational temperature was derived to be 17.83 K adopting $\tau$ = 1.24 and $R_\mathrm{(2,2)/(1,1)}$ = 0.5. The line-width of (1,1) and (2,2) profiles were assumed to be the same. We added Gaussian noise ($\sigma_\mathrm{noise} \simeq 0.04$) to these profiles and sampled 10$^5$ times. 

Figure \ref{fig:error_esti}-(c) and (d) shows histograms of the distributions of the sampled optical depth and rotational temperature, respectively. These distributions are close to Gaussian. (black lines in Figure \ref{fig:error_esti}-(c) and (d)). The standard deviation of the resultant optical depth and rotational temperatures for samples were 0.092 and 0.330 K, respectively. In this paper, we determined the error of these physical parameters were $\pm$ 0.10 and $\pm$ 0.4 K, respectively.

\bsp	
\label{lastpage}
\end{document}